\documentclass[10pt, final, journal, letterpaper, twocolumn]{IEEEtran}

\usepackage{cite}
\usepackage{amsmath,amsthm,amssymb,amsfonts}
\usepackage{algorithm}
\usepackage{algorithmic}
\usepackage{graphicx}
\usepackage{color}
\usepackage[normalsize]{subfigure}

\newtheorem{remark}{Remark}

\begin{document}

\title{Wireless Information and Power Transfer for Multi-Relay Assisted Cooperative Communication}
\author{Yuan~Liu,~\IEEEmembership{Member,~IEEE}
%\thanks{Manuscript received 15 November, 2015; revised 14 January, 2016; accepted 11 February, 2016. This work was supported in part by the Natural Science Foundation of China under Grant 61401159, in part by the Open Research Fund of the National Mobile Communications Research Laboratory, Southeast University, Nanjing, China, under Grant 2016D06, and in part by the Fundamental Research Funds for the Central Universities. The associate editor coordinating the review of this manuscript and approving it for publication was X. Chen.}
\thanks{Y. Liu is with the School of Electronic and Information Engineering, South China University of Technology, Guangzhou 510641, China (e-mail: eeyliu@scut.edu.cn).}
}

\maketitle

\vspace{-1.5cm}

\begin{abstract}
  In this paper, we consider simultaneous wireless information and power transfer (SWIPT) in multi-relay assisted two-hop relay system, where multiple relay nodes simultaneously assist the transmission from source to destination using the concept of distributed space-time coding. Each relay applies power splitting protocol to coordinate the received signal energy for information decoding and energy harvesting. The optimization problems of power splitting ratios at the relays are formulated for both decode-and-forward (DF) and amplify-and-forward (AF) relaying protocols. Efficient algorithms are proposed to find the optimal solutions. Simulations verify the effectiveness of the proposed schemes.
\end{abstract}


\section{Introduction}

Energy harvesting has emerged as a promising approach to prolong lifetime of energy constrained wireless networks. However, the traditional energy harvesting technologies rely on natural energy sources (like solar, wind, vibration and etc.), which cannot be controlled and are not always available. Recently, radio-frequency (RF) signals radiated by ambient transmitters becomes a viable new source for wireless energy harvesting. Since RF signals carry energy as well as information at the same time, simultaneous wireless information and power transfer (SWIPT) enables efficient resource allocation at transceiver designs and thus has drawn a significant attention in wireless communications.

Two practical schemes for wireless information and power transfer were proposed in \cite{ZhangHo} and \cite{ZhouZhang}, namely ``time switching"
where the receiver switches between decoding information and harvesting energy, and ``power splitting" where the receiver splits the signal power into two parts for decoding information and harvesting energy. The authors in \cite{LiuZhang1} and \cite{LiuZhang2} investigated capacity-energy tradeoff for time switching protocol and the power splitting protocol, respectively.
SWIPT has been studied for various aspects, such as energy efficiency \cite{Ng2} and physical-layer security \cite{Ng1,ZhangTIFS,MengTWC}.
%%
%Channel estimation for SWIPT was studied in \cite{Zeng2}.
%%
%Energy efficiency of SWIPT was investigated in \cite{Ng2}.
%%
%SWIPT was also applied to enhance physical-layer security (e.g., \cite{Ng1}).

In cooperative or sensor networks, the intermediate relay nodes often have less battery storage and require external charging to remain active. Thus energy harvesting is more important for relay or sensor nodes. A handful of works studied SWIPT in cooperative relay systems. For instance, the outage probability and the ergodic capacity for time switching and power splitting protocols were derived in \cite{Nasir}, where the relay harvests a fraction of energy from the source, and the relay uses the harvested energy to forward the source's information to the destination.
%
%The authors in \cite{Gurakan} considered a multiuser scenario where each user shares a portion of the harvested energy with other users, and the optimization of energy arrivals for throughput maximization using Lagrangian duality was proposed. Note that \cite{Gurakan} considered a full-duplex relay which can transmit and receive data/energy simultaneously.
%
In \cite{Ding}, the authors investigated wireless information and power transfer in cooperative networks where the randomly located relays assist one source-destination pair, and outage probability and diversity gain were characterized by stochastic geometry. The authors in \cite{Ding2} studied the outage performance of power strategies at an energy harvesting relay for multiple source-destination pairs. The optimal time-switching ratio was investigated in full-duplex relaying system under different communication modes in \cite{Zhong}. Power splitting for interference relay channels using game theory was proposed in \cite{ChenGame2015}. The authors in \cite{Quanzhong2014} studied secure relay beamforming design with SWIPT in two-hop non-regenerative relay systems.
Efficient suboptimal algorithms for SWIPT in multiple-input multiple-output orthogonal frequency-division multiplexing
(MIMO-OFDM) non-regenerative relaying were proposed in \cite{Xiong2015}.
Protocols and optimization for SWIPT based OFDM relaying were investigated in \cite{YuanVT}. Self-energy recycling for wireless-powered relay was studied in \cite{Zeng1}. In \cite{Gong}, the authors studied the time switching protocol in multi-relay network, and joint energy harvesting and beamforming schemes were proposed.

In view of the related works on SWIPT in relay systems, it is found that most of these works assume that only one relay assists the source-destination transmission.
Though the work \cite{Ding} considered multiple relays,
the authors focused on outage analysis with fixed power splitting coefficients. As the RF signals radiated by the source can be received by all relays, the multiple relays can simultaneously coordinate the received energy and information using the same signals from the source. Therefore, without extra source power, the system performance can be greatly improved by utilizing the potential of spatial diversity of the distributed relay nodes. This motivates our study.

In this paper, we consider SWIPT in two-hop relaying systems, where the multiple relays assist the transmission from the source to destination. The source node is with fixed energy supply, and the relay nodes have no energy or are not willing to expend their own energy to help the source. Assuming power splitting applied at the relay receivers, the relays split a portion of the received energy as the source of power to simultaneously forward the signals of the source by using distributed
space-time codes \cite{Laneman1,Laneman2,Yindi,Yiu}.
%To our best knowledge, this is the first work on studying SWIPT in multi-relay assisted cooperative communication.
Both decode-and-forward (DF) and amplify-and-forward (AF) relaying protocols are considered. Our goal is to maximize the end-to-end rate by optimizing the relays' power splitting ratios. We find the optimal solutions by efficient algorithms.

The remainder of the paper is organized as follows: Section II describes the system model. Problem formulations and proposed algorithms are detailed in Section III. Simulations results are provided in Section IV. Finally we conclude the paper in Section V.

\section{System Model}

\begin{figure}[t]
\begin{centering}
\includegraphics[scale=0.8]{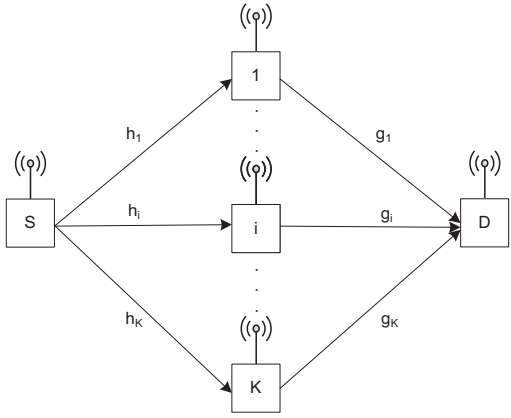}
\vspace{-0.1cm}
 \caption{System model.}\label{fig:system}
\end{centering}
\vspace{-0.3cm}
\end{figure}

We consider a two-hop relay  network model as shown in Fig.~\ref{fig:system}, where multiple relays assist the transmission from the source $\textsf{S}$ to the destination $\textsf{D}$. All nodes are equipped with a single omnidirectional antenna. The relay set is denoted as $\mathcal R=\{1,\cdots,K\}$.
We assume that the direct link between $\textsf{S}$ and $\textsf{D}$ is blocked due to the shielding effect caused by obstacles, which is a well-known Type-II relay model in 3GPP LTE-A. Such that the transmit signals from $\textsf{S}$ need to be forwarded by the relay set $\mathcal R$. Denote $h_i$ as the channel gain between $\textsf{S}$ and relay $i$, and $g_i$ as the channel gain between relay $i$ and $\textsf{D}$. Moreover,  we assume that the additive white Gaussian noises (AWGN) at all nodes are independent circular symmetric complex Gaussian random variables, each having zero mean and unit variance. The channel fading is modeled by large-scale path loss and small-scale Rayleigh fading. The transmission from the source to destination is divided into consecutive frames. It is assumed that the channel fading remains unchanged within each transmission frame but varies from one frame to another. We also assume that perfect channel state information (CSI) are available at all nodes.

Each relay node has the energy harvesting function to harvest energy from the received signals by power splitting. The source $\textsf{S}$ is with a fixed energy supply, while each relay has no energy (or is not willing to expend its own energy) to help the source, but they can forward the source's information by using the energy harvested from the source. More specifically, each relay $i$ splits a portion of the received signal energy $\alpha_i$ for information decoding, and the rest part $1-\alpha_i$ for energy harvesting. The harvested energy at each relay is used as the transmit power for forwarding the source's information.
The relay nodes operate on half-duplex mode so that the two-hop transmission is divided into two phases. In the first phase, $\textsf{S}$ broadcasts the signals to the relays. In the second phase, all relays form a distributed MIMO system and simultaneously forward the signals to the destination by using distributed space-time codes.

%In this paper, we assume that energy loss and nonlinearity, due to processing and circuitry at the transmit/receive, are neglected for simplicity.
%The assumption is widely adopted in the prior works.

\section{Optimization Framework}

In this section, we detail the problem formulations and optimal solutions for both DF and AF, respectively.

\subsection{Decode-and-Forward Relaying}

Let $P$ denote the transmit power of the source. For each relay node $i$, the received signal energy is $Ph_i$. By power splitting with the ratio $\alpha_i$, a portion of $Ph_i\alpha_i$ is used to information decoding and the rest $Ph_i(1-\alpha_i)$ is input into energy harvester as the relay power. For multi-relay assisted DF protocol, the signal-to-noise ratio (SNR) of the first and second hops are given by \cite{Laneman1,Laneman2}, respectively

\begin{equation}
\textsf{SNR}_{DF,1}=\min_{i\in\mathcal R}Ph_i\alpha_i,
\end{equation}
and
\begin{equation}
\textsf{SNR}_{DF,2}=\sum_{i\in\mathcal R}\zeta Ph_i(1-\alpha_i)g_i,
\end{equation}
where $\zeta$ is the energy conversion efficiency.
It is notable that in traditional DF scheme without power splitting (or $\alpha_i=1$), the source transmit rate is limited by the worst channel gain $h_i$ among the relay set $\mathcal R$ in the first hop. However, by considering power splitting at relays, the source transmit rate is limited by the minimum received SNR among the relay set, which significantly complicates the problem and makes this study nontrivial.

To ensure the reliable recovery of the transmit signal at every relay in $\mathcal R$, the source transmit rate should not be higher than $\log_2(1+\textsf{SNR}_{DF,1})$. Moreover, the source transmit rate should not be higher than $\log_2(1+\textsf{SNR}_{DF,2})$ to ensure reliable signal decoding at the destination. Therefore, the end-to-end rate should be \cite{Laneman1,Laneman2}
\begin{equation}\label{eqn:rdf}
R_{DF} \leq \frac{1}{2}\min\left\{\log_2(1+\textsf{SNR}_{DF,1}),\log_2(1+\textsf{SNR}_{DF,2})\right\}.
\end{equation}
Here the factor $\frac{1}{2}$ is due to the cooperative transmission taking two time slots.

Our goal is to maximize the end-to-end rate by optimizing the power splitting ratios of the relays. The problem can be formulated as
\begin{subequations}
\begin{align}
\textbf{P1:}~~~  \max_{\{\alpha_i\}}~~&R_{DF}\\
  s.t.~~~~&0\leq\alpha_i\leq1,~\forall i\in\mathcal R\label{eqn:cons1}.
\end{align}
\end{subequations}

Note that in \eqref{eqn:rdf} both $R_{DF}\leq\frac{1}{2}\log_2(1+\textsf{SNR}_{DF,1})$ and $R_{DF}\leq\frac{1}{2}\log_2(1+\textsf{SNR}_{DF,2})$ are convex sets, and so is their intersection. In addition, the constraint \eqref{eqn:cons1} is affair. Therefore the problem \textbf{P1} is a convex problem.
We can find the optimal solution with closed-form by carefully exploring the structure of the rate expression. In the following we detail the derivation.

First, for the DF case, the optimal solution must happen at $\textsf{SNR}_{DF,1}=\textsf{SNR}_{DF,2}$, which implies that the condition
\begin{equation}
  \min_{i\in\mathcal R}Ph_i\alpha_i=\sum_{i\in\mathcal R}\zeta Ph_i(1-\alpha_i)g_i
\end{equation}
should be always satisfied. To this end, we find that the following condition must hold for each $i\in\mathcal R$ at the optimum:
\begin{equation}
  Ph_i\alpha_i=\sum_{i\in\mathcal R}\zeta Ph_i(1-\alpha_i)g_i\triangleq\textsf{SNR}_{{\rm eff}},\forall i\in\mathcal R,
\end{equation}
where $\textsf{SNR}_{{\rm eff}}$ is the effective SNR and is a constant. This means that the received SNR at all relays should be the same at the optimum. Then we have
%
%\begin{equation}
%  \alpha_i=\frac{\textsf{SNR}_{{\rm eff}}}{ph_i}
%\end{equation}
%and
\begin{align}
  \textsf{SNR}_{{\rm eff}}&=\sum_{i\in\mathcal R}\zeta Ph_i\left(1-\frac{\textsf{SNR}_{{\rm eff}}}{Ph_i}\right)g_i\nonumber\\
  &=\sum_{i\in\mathcal R}\zeta Ph_ig_i-\textsf{SNR}_{{\rm eff}}\sum_{i\in\mathcal R}\zeta g_i.
\end{align}
Combining the constraint \eqref{eqn:cons1}, we get
\begin{equation}\label{eqn:eff}
  \textsf{SNR}_{{\rm eff}}=\min\left\{\frac{\sum_{j\in\mathcal R}\zeta Ph_jg_j}{1+\sum_{j\in\mathcal R}\zeta g_j},~
  \min_{j\in\mathcal R}Ph_j\right\}.
\end{equation}

%From \eqref{eqn:eff}, it is trivial that $R_{DF}=\frac{1}{2}\log_2(1+\textsf{SNR}_{{\rm eff}})$ achieves maximum when $p^*=P$.

Finally, the optimal power splitting ratio at every relay can be obtained as
\begin{equation}\label{eqn:alpha-opt-df}
  \alpha_i^*=\frac{\textsf{SNR}_{{\rm eff}}}{Ph_i}=\min\left\{\frac{\sum_{j\in\mathcal R}\zeta h_jg_j}{h_i(1+\sum_{j\in\mathcal R}\zeta g_j)},~\frac{\min_{j\in\mathcal R}h_j}{h_i}\right\}.
\end{equation}

\subsection{Amplify-and-Forward Relaying}

For the AF case, the rate expression can be easily obtained as \cite{Laneman1,Laneman2}
\begin{align}
R_{AF}&=\frac{1}{2}\log_2\left(1+\sum_{i\in\mathcal R}\frac{\zeta Ph_i\alpha_iPh_i(1-\alpha_i)g_i}{1+Ph_i\alpha_i+\zeta Ph_i(1-\alpha_i)g_i}\right)\nonumber\\
&=\frac{1}{2}\log_2\left(1+\sum_{i\in\mathcal R}\frac{\zeta P^2h_i^2\alpha_i(1-\alpha_i)g_i}{1+Ph_i\alpha_i+\zeta Ph_i(1-\alpha_i)g_i}\right).\label{eqn:raf}
\end{align}

Similar to the DF case, the problem formulation of the AF case can be expressed as
\begin{subequations}
\begin{align}
\textbf{P2:}~~~  \max_{\{\alpha_i\}}~~&R_{AF}\\
  s.t.~~~&0\leq\alpha_i\leq1,~\forall i\in\mathcal R.
\end{align}
\end{subequations}

\textbf{P2} is a nonconvex problem since the rate expression $R_{AF}$ is not concave. In the following, we find its optimal solution in an efficient way.
%To make the problem more tractable, we adopt the following upper bound
%approximation:
%%
%\begin{align}\label{eqn:appro}
%R_{AF}&\approx\frac{1}{2}\log_2\left(1+\sum_{i\in\mathcal R}\frac{p^2h_i^2\alpha_i(1-\alpha_i)g_i}{ph_i\alpha_i+ph_i(1-\alpha_i)g_i}\right)\nonumber\\
%&=\frac{1}{2}\log_2\left(1+p\sum_{i\in\mathcal R}\frac{h_i\alpha_i(1-\alpha_i)g_i}{\alpha_i+(1-\alpha_i)g_i}\right).
%\end{align}
%%
%The approximation is based on high SNR but it is proved to be tight even in low SNR regime (refer to \cite{YuanAuction} and the references therein).
%
%Using \eqref{eqn:appro}, $R_{AF}$ achieves maximum when $p^*=P$ for any given $\{\alpha_i\}$. Therefore, we just need to determine optimal $\{\alpha_i^*\}$.

First, it is observed that the summation term in \eqref{eqn:raf} is decomposable since $\{\alpha_i\}$'s are independent each other. Combining the fact that maximizing $\log_2(1+x)$ is equivalent to maximizing $x$, we can decompose the original problem into $K$ independent subproblems, each corresponding to one relay and having the identical structure. Each subproblem can be expressed as
\begin{subequations}\label{eqn:subp}
\begin{align}
\max_{\alpha_i}~~&\frac{\zeta P^2h_i^2\alpha_i(1-\alpha_i)g_i}{1+Ph_i\alpha_i+\zeta Ph_i(1-\alpha_i)g_i}\label{eqn:rho}\\
  s.t.~~~&0\leq\alpha_i\leq1.
\end{align}
\end{subequations}

The problem in \eqref{eqn:subp} is still nonconvex. However, it is readily testified that the numerator of \eqref{eqn:rho} is nonnegative and concave, while the denominator is convex and positive. Therefore, the problem in \eqref{eqn:subp} is a concave fractional programming problem \cite{Boyd}.
By defining $\phi_1(\alpha_i)\triangleq\alpha_i(1-\alpha_i)$ and $\phi_2(\alpha_i)\triangleq Ph_i\alpha_i+\zeta Ph_i(1-\alpha_i)g_i$, and introducing the new variables $x_i\triangleq\frac{\alpha_i}{\phi_2(\alpha_i)}$ and $y_i\triangleq\frac{1}{\phi_2(\alpha_i)}$, the problem in \eqref{eqn:subp} can be transformed as
\begin{subequations}\label{eqn:trans}
\begin{align}
\max_{x_i,y_i}~~&\zeta P^2h_i^2g_iy_i\phi_1\left(x_i/y_i\right)\\
  s.t.~~~&y_i\phi_2\left(x_i/y_i\right)\leq1\\
  &x_i\leq y_i\\
  &x_i\geq0,y_i>0.
\end{align}
\end{subequations}

Now the problem in \eqref{eqn:trans} is a convex problem and can be solved very efficiently using the interior-point method \cite{Boyd}. After obtaining the optimal $x_i^*$ and $y_i^*$, the optimal $\alpha_i^*$ can be recovered by letting $\alpha_i^*=x_i^*/y_i^*$.

\begin{remark}
Our optimization framework does not need a centralized controller and the optimal power splitting ratio $\alpha_i^*$ can be determined by individual relays. Specifically, for the DF case, each relay $i$ broadcasts its channel gains $h_i$ and $g_i$, thus each relay $i$ can know all channel gains $\{h_i\}$ and $\{g_i\}$ and then determine $\alpha_i^*$ using \eqref{eqn:alpha-opt-df}. For AF case, the original problem \textbf{P2} is decoupled into $K$ subproblems as \eqref{eqn:subp}, where each subproblem corresponds to one relay for finding $\alpha_i^*$ using its own channel gains $h_i$ and $g_i$. Therefore, in our paper, the optimal solutions for both DF and AF can be found in a distributed way.
\end{remark}

\section{Simulation Results}

In this section, we evaluate the proposed schemes via simulations.  We consider a two-dimensional plane where the source and destination are fixed at $(0,0)$ and $(10,0)$ m, respectively, and the relay nodes are randomly but uniformly distributed in a square with length of $2$ m whose central point is on the line between the source and destination. $K=5$ relays are located in the square. The central point of the relay square is at $d=(5,0)$, i.e., the relay region is at the middle of the source and destination.  Each fading is modeled as $c\cdot L^{-\theta}$, where $c$ is the Rayleigh fading factor, $L$ is transmitter-receiver distance, and $\theta$ is the path loss exponent which is set to be $3$. We set the energy conversion efficiency coefficient $\zeta=60\%$.
%A total of $10000$ independent channel realizations were generated, each associated with different relay locations.

We first consider the traditional best relay selection (BRS) as the performance benchmark: Assuming that the relay set $\mathcal R$ contains only one relay node, the optimal source power and power splitting ratio can be obtained using the proposed algorithms for both DF and AF; then do an exhaustive search over all relays and select a relay that results in best system performance.

%%
%\begin{figure}[t]
%\begin{centering}
%\includegraphics[scale=0.7]{setup.eps}
%\vspace{-0.1cm}
% \caption{Two-dimensional plane of node locations, where the source and destination are located at $(0,0)$ and $(10,0)$, respectively.}\label{fig:setup}
%\end{centering}
%\vspace{-0.3cm}
%\end{figure}
%%

%
\begin{figure}[t]
\begin{centering}
\includegraphics[scale=0.55]{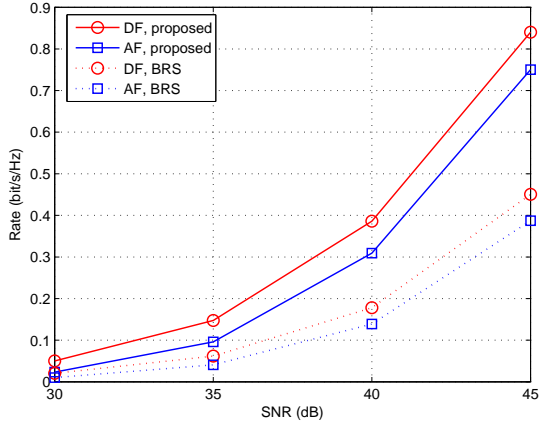}
\vspace{-0.1cm}
 \caption{Performance comparison of different schemes.}\label{fig:rate}
\end{centering}
\vspace{-0.3cm}
\end{figure}

Fig.~\ref{fig:rate} compares rate performances of different schemes. It is observed that the proposed DF and AF schemes perform closely, and the DF scheme is slightly superior to the AF scheme. Moreover, we also observe that the proposed schemes significantly outperform the traditional BRS schemes, which demonstrates the superiority of the proposed schemes. Note again that the performance gain compared to the BRS schemes is without any extra source power, since the relays use the same signals of the source. Thus the proposed schemes is preferable even in an energy-efficiency perspective.

\begin{figure}[t]
\begin{centering}
\includegraphics[scale=0.55]{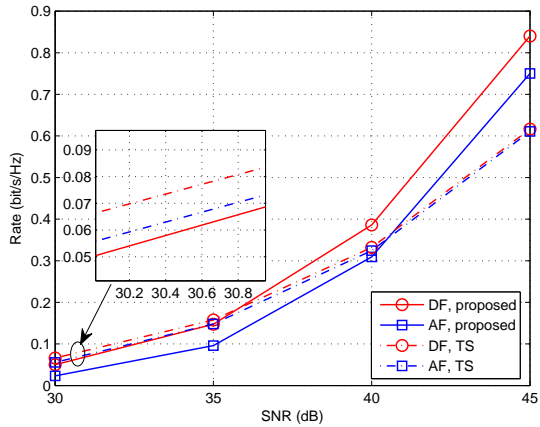}
\vspace{-0.1cm}
 \caption{Performance comparison with time switching protocol.}\label{fig:ts}
\end{centering}
\vspace{-0.3cm}
\end{figure}

We then compare the proposed power splitting schemes with time switching (TS) schemes in \cite{Gong}, where we exhaustively search the time allocation variable $t$ over the range $[0,1]$ with very small intervals for finding the optimal $t^*$ that makes the rate maximum.
We observe that the power splitting protocol outperforms the time switching  protocol in high SNR region, while the time switching  protocol is better in low SNR region. Our observation is in accord with the results in \cite{Nasir} where single-relay is considered.

\section{Conclusion}

This paper studied simultaneous wireless information and power transfer in multi-relay assisted two-hop cooperative communication, where multiple relay nodes use power splitting protocol to simultaneously coordinate the usage of the received signals for information decoding and energy harvesting.
Power splitting ratios at the relays were optimized for both DF and AF relaying strategies. Simulation results show that the proposed schemes significantly outperform the traditional best relay selection schemes. Moreover, we demonstrate that power splitting is better in high SNR region and time switching is favoured in low SNR region in relay networks.

\bibliographystyle{IEEEtran}
\bibliography{IEEEabrv,CL2016-0122-multirelay}

\end{document}